# An Experimental and Computational Study on Material Dispersion of 1-Alkyl-3-Methylimidazolium Tetrafluoroborate Ionic Liquids

Carlos Damián Rodríguez Fernández[a], Yago Arosa[a], Bilal Algnamat[a,b], Elena López Lago[a*], Raúl de la Fuente[a]

[a] *Nanomateriais, Fotónica e Materia Branda (NaFoMat), Departamento de Física Aplicada e Departamento de Física de Partículas, Universidade de Santiago de Compostela, Campus Vida, E-15782 Santiago de Compostela, Spain*

[b] *Department of Physics, College of Science, Al-Hussein Bin Talal University, Ma'an, Jordan*

*Corresponding author: elena.lopez.lago@usc.es*

## Abstract

The material dispersion of the $[C_kmim][BF_4]$ (k = 2,3,4,6,7,8,10) family of ionic liquids is measured at several temperatures over a broad spectral range from 300 nm to 1550 nm. The experimental curves are fitted to a modified three-resonance Sellmeier model to understand the effect of temperature and alkyl chain length in the dispersion. From the parameters of the fitting, we analyze the influence that the different constituents of these ionic liquids have in the dispersion behaviour. In addition, a semi-empirical approach combining simulated electronic polarizabilities and experimental densities for predicting the material dispersion is successfully tested by direct comparison with the experimental results. The limitations of this method are analyzed in terms of the structure of the ionic liquids. The results of this work aim to increase our knowledge about how the structure of an ionic liquid influences its material dispersion. Understanding this influence is fundamental to produce ionic liquids with tailored optical properties.

## 1. Introduction

Ionic Liquids (ILs) are materials that are attracting much attention in the last decades. They are formed by the combination of an organic cation and a poorly coordinated anion producing materials with low melting points that usually are liquids at room temperature. The result is a liquid composed uniquely by ions owning exclusive properties such as very low vapour pressure, low flammability, high thermal and chemical stability or a large degree of tunability. ILs are the focus of an intense research from the perspective of different fields ranging from battery design[1–4] to mechanic lubrication[5–8] or extraction science[9–11]. In the field of photonics, ILs are starting to attract interest as promising optical materials despite there is a long way to go in the study of these properties. For instance, ILs with tunable photoluminescence[12–14] or photochromism[15] were recently synthesized. Photoluminescence of a recently synthesized tailored IL was used as a detector for traces of a specific chemical contaminant[16]. There are also examples of ILs owning thermochromic behaviour arising from different molecular mechanisms[17,18]. Furthermore, ILs can be tuned to produce ionic liquid crystals which are ILs showing liquid crystal behaviour[19–22]. The recent research evidences that the wise election of the combination of cation and anion enables the production of ILs with attractive tailored properties for optical materials. In most cases, the refractive index is an essential property when designing an optical device. ILs with tuned refractive index would be useful for a large set of applications such as immersion liquids for microscopy[23] or lithography[24], as variable focus lenses[25] or as a part of more complex photonic devices. For instance, an all-optical attenuator was recently built[26] based on an optical fiber filled with an IL. The changes on the refractive index of this IL driven by the change in temperature produced by the light intensity in the fiber was used to provide a mechanism to selectively switch the transmission of light.

Pioneering measurements of refractive indices at single wavelengths, namely the sodium D line (589 nm), for different families of ILs were done by different authors[27–29]. The measurement of refractive indices of different families of ILs provides valuable information about the relationship existing between them and the molecular structure of ILs. For instance, the anions are the part of the ILs that most influences the refractive index value while slight modifications of the cations such as changing the length of the alkyl chain provides a fine tuning of it. Relations of the refractive index with other properties were also studied[30]. In parallel with experimental measurements, the refractive index of ILs was studied also from the perspective of different computational approaches since the very first moment[31–39]. These approaches are a fundamental tool for predicting and understanding the refractive index since the experimental characterization of each physically feasible IL is not possible due to the virtually infinite combinations of ions producing them. Usually, these computational approaches provide new insights on the refractive index of ILs by relating it with their structure by means of the electronic polarizability. Some of them are based on computational techniques relating experimental refractive indices with the structure of the molecules through statistical approaches such as neuronal networks[32], quantitative structure-property relationships[31], Thole models[34] or designed regression analysis[33]. Meanwhile, other methods predict the refractive index by calculating the static electronic polarizability by *ab initio* algorithms[36–39] such as density functional theory (DFT) or Møller–Plesset perturbation theory (MP).

Thanks to this battery of experimental and computational works, nowadays we have a better knowledge of how the value of refractive index at a given wavelength correlates with the structure of ILs. Unfortunately, the literature about the refractive index spectral and thermal behaviour as well as its dependence on the IL structure is much more limited. This lack of information is a direct consequence of the absence of works dealing with material dispersion both from the experimental and computational points of view. In the case of experimental works, only few authors published multi-wavelength refractive index measurements[25,40–44]. This limitation in the available experimental data has a direct influence in the number of computational studies dealing with dispersion. Furthermore, most of the purely *ab initio* works relate refractive index with the electronic polarizability. This electronic polarizability is by itself a wavelength dependent function but calculations are mainly done in the static limit[33,38]. Hence, the dependence on wavelength of the electronic polarizability is absolutely neglected and dispersion information is not preserved.

In a previous publication[44], we experimentally measured the refractive index of several families of ILs not at a discrete set of wavelengths but in a continuous fashion covering a wide spectral range from 400 nm to 1000 nm as well as their thermal response between 298 K and 323 K. We employed this experimental data for modelling different imidazolium-based ILs as a function of their structure. In order to do that, we considered a Sellmeier dispersion formula with a single resonance. Even though the simplicity of this model, the experimental material dispersion was properly described in the whole spectral and thermal ranges. The results showed some interesting features. For instance, changing the length of the alkyl chain of an imidazolium cation in combination with a fixed anion does not change the position of the effective resonance governing the dispersion behaviour while the refractive index magnitude presents a rational dependence on the number of carbons.

In this work we aim to obtain new insights on how the structure of ILs affects the material dispersion by considering a wider spectral range than in our previous publications. The study is restricted to the 1-alkyl-3-methylimidazolium tetrafluoroborate family of ILs, $[C_k mim][BF_4]$ with $k$= 2, 3, 4, 6, 7, 8 and 10. However, the approach followed in this work can be extended to the study of other families of ILs; for example 1-alkyl-3-methylimidazolium cations combined with other anions ($[C_k mim][NTf_2]$, $[C_k mim][OTf]$, ...) or families formed by ILs sharing the same cation and anions but that differ exclusively in the length of its alkyl chain such as the $[C_2 mim][C_k SO_4]$ family.

Material dispersion of the ILs was measured in the spectral range from 300 nm to 1550 nm at temperatures from 293 K to 313 K. The material dispersion was measured by Refractive Index Spectroscopy by Broadband Interferometry (RISBI), a powerful interferometric technique based on spectrally resolved white light interferometry (SRWLI)[45–47]. The material dispersion curves were fitted to a three-resonance Sellmeier model commonly used in the chromatic dispersion characterization of optical materials[48–50]. From the fitting parameters, valuable information about the influence of the ILs' structure in the material dispersion is extracted. Furthermore, the experimental measurements were complemented with DFT calculations including optical absorption spectrum and wavelength-dependent electronic polarizability. These wavelength-dependent calculations were used together with a semi-empirical model to faithfully reproduce the experimental material dispersion in both the spectral and thermal ranges. This simple semi-empirical model constitutes the first computational approach to the modelling of the material dispersion of ILs.

## 2. Experimental

### 2.1 Materials

In this paper we study the material dispersion of seven ILs of the 1-alkyl-3-methylimidazolium tetrafluoroborate [$C_k$mim][$BF_4$] family with $k$= 2, 3, 4, 6, 7, 8 and 10. The ILs were purchased from Io-Li-Tec and their water content was checked by means of a Mettler Toledo Karl-Fisher titrator coulometer. We established as acceptable an upper limit of 700 ppm of water contamination for the ILs to be used in this work. ILs presenting water impurities higher than this concentration were subjected to a drying process to fit this limit. The drying consisted on vacuum pumping the IL samples at room temperature while stirring for at least 48 hours. After completing the process, the amount of water was measured again to check the effectiveness of the water removal. In order to prevent further contact with air moisture up to the moment of the measurement, dried liquids were kept in glass vials and closed with screw caps fitted with a silicone septum to ensure their isolation. Table 1 shows the ILs selected for this work, the purity provided by the supplier and their water content, after drying when necessary.

| LI | CAS | Purity (%) | Water (ppm) |
|---|---|---|---|
| [$C_2$mim][$BF_4$] | 143314-16-3 | 99.1 | 258 |
| [$C_3$mim][$BF_4$] | 244193-48-4 | 99.1 | 343 |
| [$C_4$mim][$BF_4$] | 174501-65-6 | 99.8 | 464 |
| [$C_6$mim][$BF_4$] | 244193-50-8 | 99.9 | 240 |
| [$C_7$mim][$BF_4$] | 244193-51-9 | 99.9 | 296 |
| [$C_8$mim][$BF_4$] | 244193-52-0 | 99.8 | 238 |
| [$C_{10}$mim][$BF_4$] | 244193-56-4 | 99.8 | 600 |

Table 1: List of ILs employed in this work and their CAS number, purity provided by supplier and water content.

### 2.2 Refractive index

The refractive index of the samples was measured at the sodium D line, using an Atago DR-M2 Multi-Wavelength Abbe Refractometer. The refractometer was calibrated with deionised water and its uncertainty at the D line is $2 \cdot 10^{-4}$. This line was selected to be the reference point for the chromatic dispersion retrieval by RISBI. Refractive indices were measured at different temperatures covering from 293 K to 313 K in steps of 2 K. The temperature was controlled by using a circulating water bath within a resolution of 0.1 K.

### 2.3 *Material dispersion*

The spectral variation of the refractive index was measured by RISBI, a white light spectral interferometry-based technique. The experimental device is constituted by two homemade instruments. The first one covers the range from 400 to 1550 nm and it is composed by a stabilized halogen light source, a Michelson interferometer and two fiber coupled spectrometers, one for wavelengths shorter than 1000 nm and the other for wavelengths longer than 900 nm. The second instrument consists of a deuterium light source, another Michelson interferometer and a prism spectrometer. The working range is 255-500 nm. Further and detailed characteristics of the instruments can be found in previous publications[46,47,51]. Material dispersion was measured at the same temperatures that those of the refractive indices by Abbe Refractometry, from 293 K to 313 K in 2 K steps. The temperature was controlled by using a circulating water bath and measured with a resolution of 0.1 K by means of a thermocouple sensor whose probe was directly introduced in the cell containing the IL sample. The experimental resolution of this device is $2 \cdot 10^{-4}$.[47]

### 2.4 *Density*

Density was measured with an Anton Paar DSA-5000 M vibrating tube density and sound velocity meter at the same range of temperatures of the refractive index measurements. The apparatus was calibrated by measuring the density of bi-distilled water and dry air at atmospheric pressure and its experimental resolution is $2 \cdot 10^{-6}$ g·cm$^{-3}$. The experimental densities of the ILs considered in this work are shown in the Supporting Information Table 1.

### 2.5 *Computational Details*

Simulations were made by DFT at the level of theory B3LYP/6-311++G(d,p). Geometry optimization for each compound was carried out over isolated ionic pairs. The stability of each reached configuration was checked through a vibrational analysis. Absorption spectra of the different ILs were obtained by means of a standard TD-DFT calculation. The theoretical frequencies and strengths of the absorption resonances were broadened by convolution with a Gaussian distribution to produce a standard absorption spectrum. Wavelength dependent electronic polarizability $\alpha(\lambda)$ was obtained by means of a Coupled Perturbed Kohn-Sham (CPKS) calculation. Electronic polarizability was simulated at discrete wavelengths from 300 nm to 1500 nm every 100 nm, a similar spectral range as the used in the material dispersion measurements.

## 3. Experimental results and discussion

As previously mentioned in Section 2, material dispersion was measured for the $[C_k mim][BF_4]$ family of ILs with $k=$ 2, 3, 4, 6, 7, 8, and 10 at a set of temperatures covering from 293 K to 313 K each 2 K and at a broad spectral range from 300 nm to 1550 nm. Figure 1 shows the material dispersion curves for the seven ILs at 303 K, reproducing the wavelength dependent refractive index in the spectral interval from 300 to 1550nm..

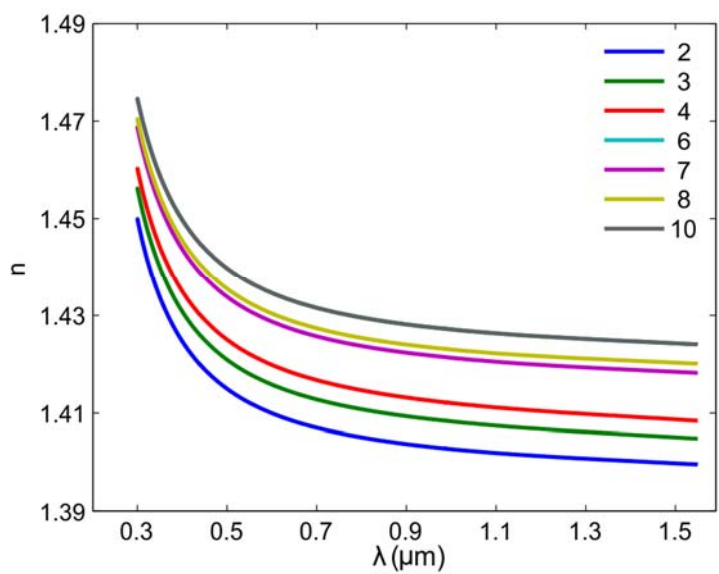

Figure 1. Experimental material dispersion of $[C_k mim][BF_4]$ ILs with $k=$ 2, 3, 4, 6, 7, 8, and 10 in the range from 300 nm to 1550 nm at T=303 K.

The qualitative behaviour of the material dispersion shows a normal dispersion regime where the refractive index grows as the wavelength decreases, with greater variations at shorter wavelengths. Regarding the magnitude of the refractive index at a specific wavelength, it grows with the alkyl chain length. The refractive index dependence on the length of the alkyl chain of the imidazolium cation was pointed out by other authors for this and other families of ILs[33]. Regarding the magnitude of the dispersion, the maximum variation of the refractive index due to the chromatism is $5.2 \cdot 10^{-2}$. In the Supporting Information Tables 2, 3 and 4, we give the value of refractive indices at selected wavelengths and temperatures for all the ILs.

Checking the agreement of our measurements with experimental data published by other authors is not possible due to the lack of available material dispersion measurements with the exception of very punctual publications[25,40–44]. However, a good validation can be done if we restrict our analysis to the refractive index at the D line, $n_D$, wavelength at which most of the experimental measurements were done. Figure 2 shows a histogram of the relative deviation of the $n_D$ measured in this work with respect to the previously published data. The comparison set includes more than 130 values from the studied ILs at several temperatures[27,29,39,52-79] measured by 30 different authors. In addition, the histogram was fitted to a Gaussian distribution without taking into account the outlier values pointed out in the figure. The obtained Gaussian distribution is centred in $\mu = 1.37 \cdot 10^{-4}$ and shows a standard deviation of $\sigma_{exp} = 2\sigma = 1.14 \cdot 10^{-3}$. Our mean relative deviation with respect the bibliography $\mu$ is below our experimental resolution which provides an important evidence of the accuracy of our measurements. On the other hand, the standard deviation of the distribution $2\sigma$ provides a striking evidence of the great dispersion of the refractive index values available in the bibliography. In fact, this standard deviation is much larger than the experimental resolution claimed for the refractive index measurements in most of the publications. We interpret this high standard deviation values as a consequence of differences in the material measured in different laboratories and not to problems in the measurement itself. Halide contaminations as well as degradation make ILs that should be transparent to look yellowish[80], increasing the refractive index value at the D line. Moreover, water contamination is also a trouble when working with ILs as it greatly decreases the value of refractive index and it is hard to avoid due to the velocity of atmospheric water absorption of most ILs[40]. Being aware of these problems, in this work we ensured the transparency of the ILs by measuring its absorption spectra (not shown) as well as minimizing the amount of water in the samples by measuring them just after drying and inside sealed cuvettes when the compatibility with the technique allowed it.

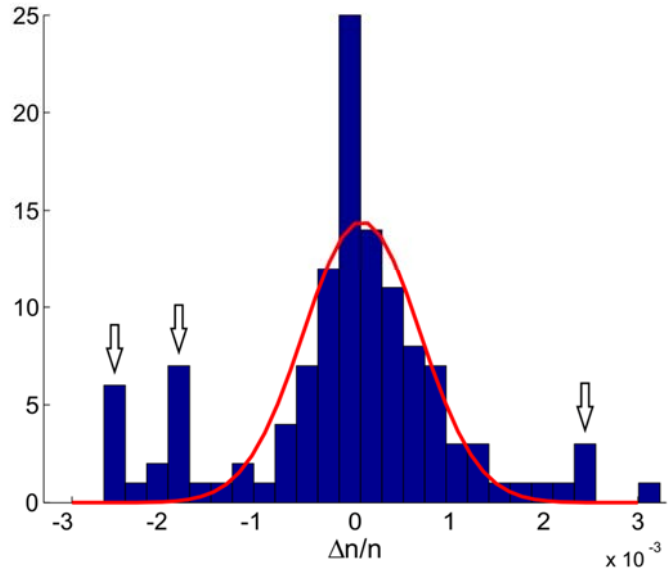

Figure 2. Histogram of the relative deviations of the $n_D$ measured in this work compared to bibliography and fitting of the data to a Gaussian distribution. The outliers marked with arrows were not taken into account for the fitting.

In this work, the material dispersion was measured in an enlarged spectral range in comparison with our previous publications[44,81] or the ones by other authors[41]. In these previous studies, the dispersion behaviour was successfully described by a one-resonance Sellmeier model. However, this approximation only holds for restricted spectral ranges and it is not valid for the spectral range measured here. In Figure 3, the material dispersion of [C4mim][BF4] at 303 K measured in this work is compared with the dispersion behaviour predicted by the one-resonance Sellmeier models of our previous work[44] and the proposed by Wu [41]. Both models describe correctly the material dispersion of the material in the range from 0.4 μm, to 1 μm. There, the deviations between the experimental data and the models are kept below $10^{-3}$. Nevertheless, at shorter and longer wavelengths the models clearly diverge from the experimental data and the registered deviations are as high as $3 \cdot 10^{-3}$. Such large deviations are not negligible and a more adequate model must be used for this extended spectral range.

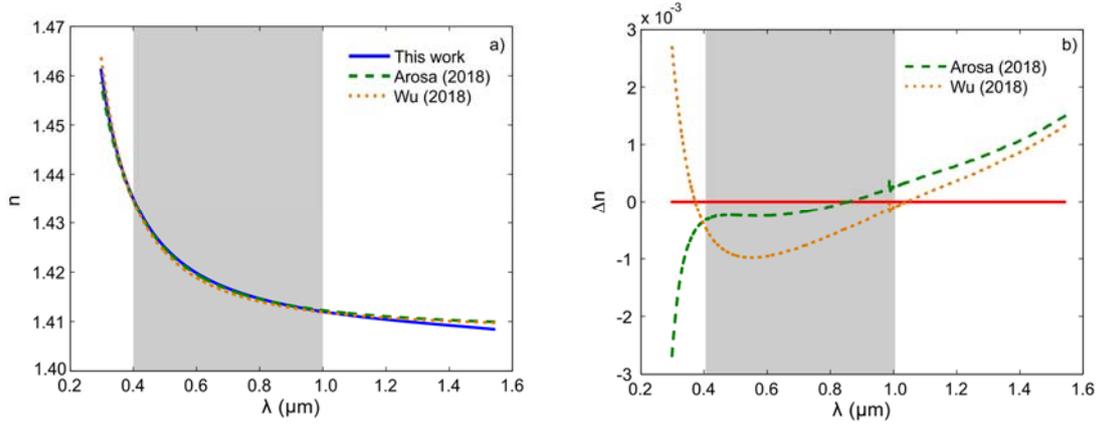

Figure 3. a) Experimental material dispersion for [C$_4$mim][BF$_4$] at 303 K from this work and comparison with the predicted by the one-resonance Sellmeier models proposed by Wu[41] and our previous work, Arosa[44]. The grey area marks the validity of the spectral range of these models. b) Absolute deviations of each model to the experimental data.

So, we considered other models that describe the new information contained in the material dispersion curves in such a broad spectral range. After performing extensive numerical simulation, we conclude that a modified three-resonance Sellmeier model provides both accuracy in the dispersion description and a good interpretation of its relation with the structure of the ILs. A standard three-resonance Sellmeier model has the following analytical expression:

$$n^2(\lambda, k, T) - 1 = \sum_{i=1}^{3} \frac{c_i(k,T)\,\lambda^2}{\lambda^2 - \lambda_i^2(k,T)} \tag{1}$$

In this equation $\lambda_i^2(k,T)$ represents a resonance wavelength and $c_i(k,T)$ its associated strength, both depending on the alkyl chain length of the liquid, $k$, and the temperature $T$. In order to validate this expression, an initial rational fit was carried out over each liquid and at each temperature with the purpose of obtaining seed values of $\lambda_i^2(k,T)$ and $c_i(k,T)$ for performing a more sophisticated nonlinear fitting of our data to equation 1. From this first approach, we discover that the position of the resonances $\lambda_i$ is almost temperature and alkyl-chain length independent. Two of them are placed in the UV region while the third one is located in the IR, very far from our measurement region, $\lambda_{IR} \sim 50\ \mu m >> \lambda$. In the event of a resonance placed very far and at larger wavelengths than the measuring range, the following approximation can be taken:

$$\frac{c_i \lambda^2}{\lambda^2 - \lambda_i^2} \cong -\frac{c_i}{\lambda_i^2}\lambda^2 = d_i \lambda^2 \tag{2}$$

Introducing equation 2 into equation 1 and taking into account that the position of the resonances is temperature and alkyl chain length independent, we obtain the modified three-resonance Sellmeier model used in this work:

$$n^2(\lambda, k, T) - 1 = \frac{c_1(k,T)\,\lambda^2}{\lambda^2 - \lambda_1^2} + \frac{c_2(k,T)\,\lambda^2}{\lambda^2 - \lambda_2^2} + d_3(k,T)\lambda^2 \tag{3}$$

We would like to discuss separately the effect that temperature and alkyl chain length have in the dispersion of our ILs on the basis of our proposed model. The temperature dependence is going to be analyzed directly on expression 3 but the alkyl chain length dependence is going to be studied in terms of the molar refraction dispersion, $R(\lambda, k)$, using a more convenient form of the dispersion established by equation 3.

We are going to start with the analysis of the temperature influence on the material dispersion. The experimental data of each IL is separately fitted to equation 3 by means of a two-step procedure. First, the experimental curves are fitted to equation 3 without imposing any constraint. In consequence, a pair of values $\lambda_1$ and $\lambda_2$ are obtained for each liquid and temperature. Afterwards, a second fitting is implemented constraining the values of $\lambda_1$ and $\lambda_2$ to be constant and using the mean values from the previous fit as a seed. For the seven ILs, the temperature dependence of the strength of the resonances was found to be similar. In Figure 4 the evolution of the resonance strength with temperature is shown for the [C$_{10}$mim][BF$_4$]. The temperature is expressed with respect to the centre of our measuring temperature interval, $T_0$=303 K.

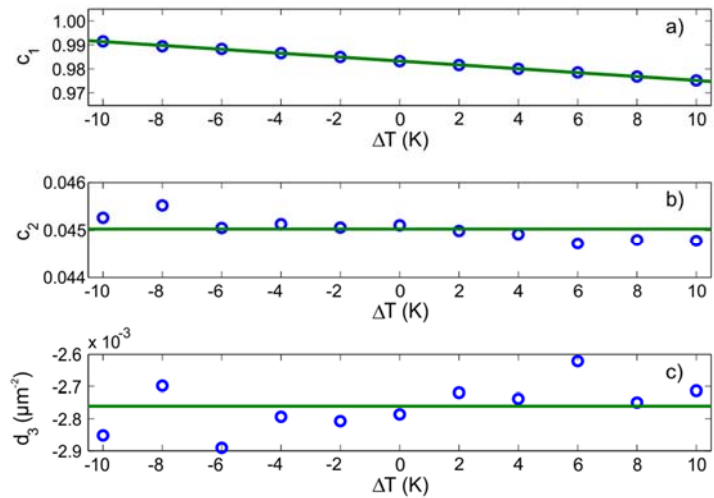

Figure 4. Temperature dependence of the resonance strengths of [C$_{10}$mim][BF$_4$] with respect to the centre of our measuring interval, T$_0$=303 K. a) c$_1$ and a linear regression on ΔT, b) c$_2$ and its average value and, c) d$_3$ and its average value.

For all the compounds, the $c_1$ coefficient clearly presents a linear dependence on temperature. The slope of this coefficient is negative and, in accordance with previous publications[44], it is tightly related to the thermo-optical coefficient (TOC) of the ILs. On the other hand, the rest of the coefficients do not show any evident trend with temperature. Even though they could admit a temperature dependent polynomial fitting, the contribution of their temperature variation to the refractive index is lower than our experimental resolution 2 10$^{-4}$. Hence, we chose to make a new fit for each IL in which we force $c_2$ and $c_3$ to be temperature independent. In this new fit the only temperature dependent variable is $c_1$ that depends linearly on temperature. Hence, the $c_1$ coefficient is split into two terms inside the Sellmeier model $c_1 = c_1' + c_1'' \cdot \Delta T$ and the temperature is again referenced to $T_0$=303 K, the centre of our measurement interval:

$$n^2(\lambda, T) - 1 = \frac{(c_1' + c_1'' \Delta T)\,\lambda^2}{\lambda^2 - \lambda_1^2} + \frac{c_2\,\lambda^2}{\lambda^2 - \lambda_2^2} + d_3\lambda^2 \qquad (4)$$

The material dispersion of each IL is described at all the temperatures by the Sellmeier model given by equation 4. The values of these parameters are presented in Table 2.

| IL | $\lambda_1$ (µm) | $c_1'$ | $c_1''$ ($10^{-4}$ K$^{-1}$) | $\lambda_2$ (µm) | $c_2$ ($10^{-2}$) | $d_3$ ($10^{-3}$ µm$^{-2}$) | $\sigma$ ($10^{-5}$) |
|---|---|---|---|---|---|---|---|
| [C$_2$mim][BF$_4$] | | 0.9079 | -71.0 | | 5.09 | -1.7 | 4.7 |
| [C$_3$mim][BF$_4$] | | 0.9234 | -76.2 | | 5.17 | -2.3 | 6.4 |
| [C$_4$mim][BF$_4$] | | 0.9356 | -73.5 | | 5.10 | -3.2 | 4.8 |
| [C$_6$mim][BF$_4$] | 0.0944 | 0.9556 | -79.5 | 0.2033 | 4.89 | -2.3 | 4.7 |
| [C$_7$mim][BF$_4$] | | 0.9641 | -78.6 | | 4.71 | -1.9 | 5.8 |
| [C$_8$mim][BF$_4$] | | 0.9684 | -81.4 | | 4.73 | -1.5 | 4.2 |
| [C$_{10}$mim][BF$_4$] | | 0.9821 | -84.6 | | 4.57 | -1.9 | 4.7 |

Table 2. Parameters and standard deviation of the fitting of the material dispersion of each IL to the temperature dependent Sellmeier model given by equation 4.

From the value of the parameters in Table 2, the contribution of the different terms in equation 4 to the refractive index can be analyzed. The major contribution comes from the lower UV resonance. The relative contribution of the second UV resonance is smaller than the 10%, while the contribution of the IR resonance is residual, less than 1%. The most relevant feature in Table 2 is that the coefficient $c_1'$ increase with the cation alkyl chain length on the same fashion as the refractive index at a fixed wavelength does. This behaviour can be observed in Figure 5 where $c_1'$ and $n_D$ are shown as a function on the alkyl chain length of the different ILs. The asymptotic trend that both magnitudes are showing was already pointed by previous authors[33,40] and analysed in our previous work [44].

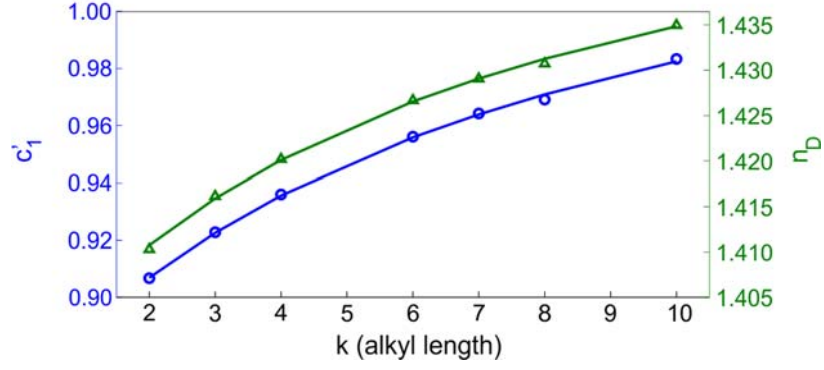

Figure 5. Refractive index at the D line, $n_D$, and the fitting coefficient $c_1'$ as a function of the alkyl chain length of the IL.

Now we are going to study the dispersion as a function of the alkyl chain length. However, we do not consider the refractive index for this analysis but the molar refraction, $R$. The molar refraction is related to the refractive index by means of the Lorentz-Lorenz equation:

$$R(\lambda, k, T) = V(k, T) \frac{n^2(\lambda, k, T) - 1}{n^2(\lambda, k, T) + 2}$$ (5)

In that expression, $V$ is the molar volume of the IL. On the other hand, the molar refraction is a magnitude proportional to the electronic polarizability, $\alpha$, by the relationship $R = \alpha N_A / 3\varepsilon_0$ being $N_A$ the Avogadro number and $\varepsilon_0$ the vacuum electrical permittivity. The variation of $R$ with temperature can be easily calculated by applying natural logarithms to equation 5 and taking the derivative of that expression with respect T:

$$\frac{1}{R} \frac{dR}{dT} = \frac{1}{V} \frac{dV}{dT} + \frac{6n}{(n^2 - 1)(n^2 + 2)} \frac{dn}{dT}$$ (6)

The first contribution corresponds to the thermal expansion coefficient while the second is related to the thermo-optic coefficient. From the experimental data, it can be deduced that both contributions are of the same order of magnitude so, for the range of temperature considered in this work, $\delta T = T_{max} - T_{min}$, it is verified:

$$\frac{\delta R}{R} = \frac{\delta T}{R} \cdot \frac{dR}{dT} \ll 1$$ (7)

Being $\delta R \cong R(T_{max}) - R(T_{min})$. Henceforth, we can consider the molar refraction as a temperature independent magnitude within our experimental resolution, $R = R(\lambda, k)$. A similar conclusion was obtained in[82] for ionic compounds with low melting points, which is the case of ILs. In consequence, studying the molar refraction dispersion instead of the refractive index dispersion provides a way to carrying out a totally temperature independent analysis of the influence of the alkyl chain length. We assume that the molar refraction obeys a functional form similar to the one used for the squared refractive index. In consequence, we use the same three-resonance Sellmeier dispersion formula expression as in equation 3 rewritten in terms of the molar refraction:

$$R(\lambda) = \frac{c_1\,\lambda^2}{\lambda^2 - \lambda_1^2} + \frac{c_2\,\lambda^2}{\lambda^2 - \lambda_2^2} + d_3\lambda^2 \qquad (8)$$

Experimental molar refraction dispersion, $R(\lambda, k)$, can be obtained from the experimental material dispersion by using the molar volumes obtained from the experimental densities of each IL and exploiting equation 6. In the following, the analysis is carried out at the centre of our temperature interval, the reference temperature $T_0$=303 K. The first step to understand the molar refraction dispersion dependence on the alkyl chain is analyzing it at specific wavelengths. As Figure 6.a) shows, this dependence is approximately linear, independently of the chosen wavelength. Furthermore, the molar volume $V$ dependence on the alkyl chain is also linear, as shown in Figure 6.b).

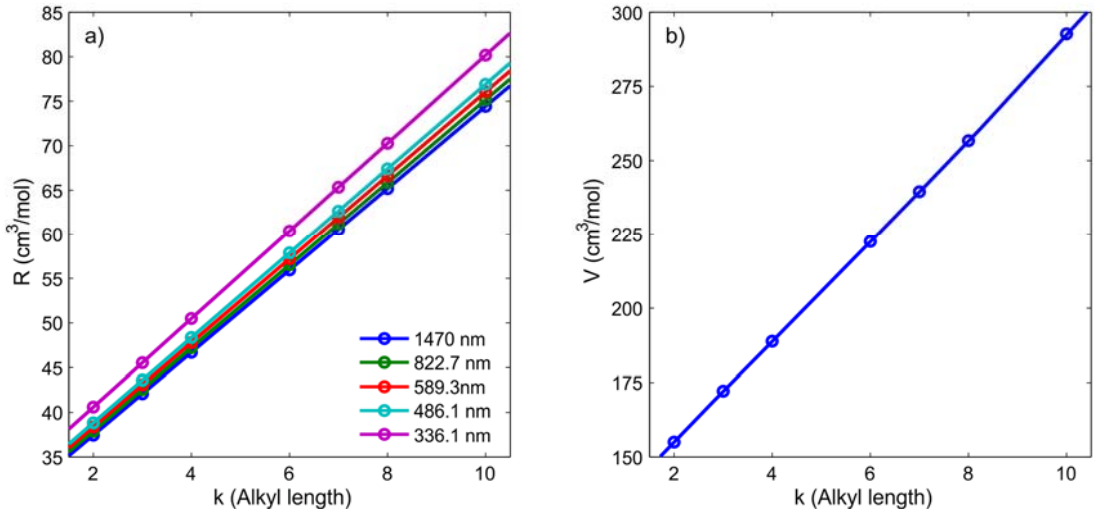

Figure 6. a) Molar refraction $R$ at different spectral lines and b) molar volume $V$ both as a function of the alkyl length.

The linear dependence of $R$ and $V$ on the alkyl chain length leads to the following expressions for these magnitudes as a function of the number of carbons in the alkyl chain, $k$:

$$V(k, T) = V_0(T) + k \cdot V_{alkyl}(T) \qquad (9)$$

$$R(k, \lambda) = R_0(\lambda) + k \cdot R_{alkyl}(\lambda) \qquad (10)$$

Equation 9 and equation 10 have a very straightforward interpretation. For the case of $V(k, T)$, there is a minimum volume of our family of ILs, $V_0$ which corresponds to contribution of the 1-methylimidazolium tetrafluoroborate, we note this compound as [mim][BF$_4$]. This minimum volume increases by an amount $V_k$ for each CH$_2$ unit attached to the [mim]$^+$ cation. It is worth to note that this linear growing of $V$ with the alkyl chain length implies that the imidazolium cation can be considered as a cylinder that enlarges its length but not its basis with each CH$_2$. Regarding $R(k, \lambda)$, its linear growing with the alkyl chain length is related to the behaviour of the electronic polarizability. It is well known that the total electronic polarizability of a compound approximately increases linearly with the number of atoms $N$ of each element in its structure $\alpha_{total} \approx \sum_i^N \alpha_i$ [30,33]. In this case, we have a minimum molar refraction $R_0$ which is again originated by the basic unit [mim][BF$_4$]. As the alkyl chain grows, the polarizability of the CH$_2$ groups is added to the total polarizability of the compound introducing an increase of $R_{alkyl}$ in the molar refraction per CH$_2$ group. The linear growing of both $V(k)$ and $R(k, \lambda)$ with the alkyl chain length predicts through the equation 6 that there are both an upper ($k \rightarrow \infty$) and a lower ($k \rightarrow 0$) limit for the refractive index of whatever IL family whose difference between members is the length of a carbon alkyl chain. This behaviour of refractive with the alkyl chain length of ILs was already noted in previous studies[33].

In order to model the $R(\lambda, k)$ behaviour along the spectrum, we split equation 5 in two different expressions, one depending on $R_0(\lambda)$ and the other depending on $R_{alkyl}$, and we assume that both, $R_0(\lambda)$ and $R_{alkyl}$, verify a dispersion equation similar to that given by equation 8. On this way, we can separately analyze the influence in the dispersion of both the [mim][BF$_4$] and the alkyl chain length.

The experimental dispersions $R_0(\lambda)$ and $R_{alkyl}(\lambda)$ are obtained from fitting $R(\lambda, k)$ at each wavelength to a linear function with respect the alkyl chain length, $k$. Afterwards, both $R_0(\lambda)$ and $R_{alkyl}(\lambda)$ are separately fitted to equation 8. Figure 7 shows the experimental dispersions $R_0(\lambda)$ and $R_{alkyl}(\lambda)$ as well as an inset with the residuals from the fitting of these curves to the proposed expressions. The Table 3 contains the parameters resulting from these fits.

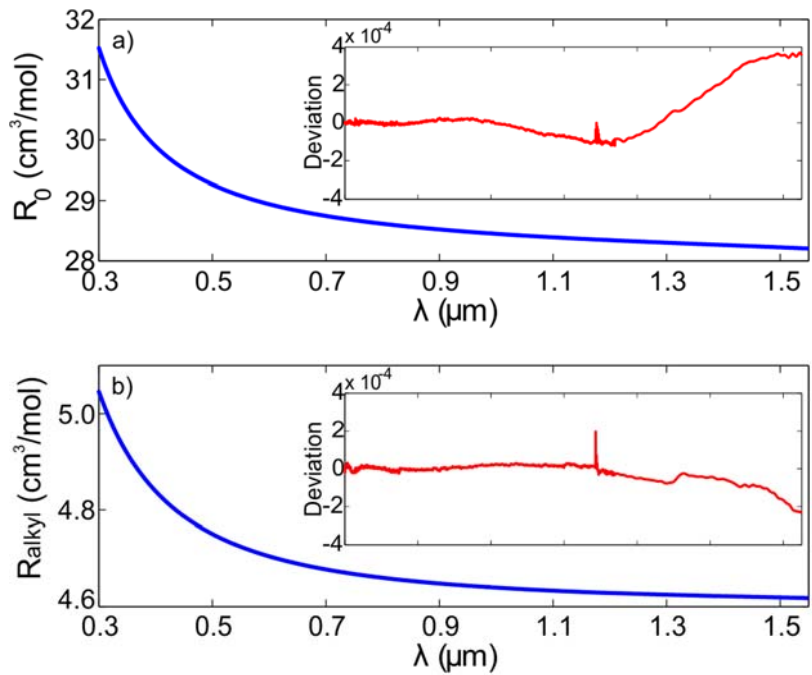

Figure 7. Molar refraction dispersion of a) $R_0(\lambda)$ and b) $R_{\text{alkyl}}(\lambda)$. The insets show in each case the absolute deviation of the experimental curves from the dispersion described by equation 8.

| $R_0(\lambda)$ | | | | |
|---|---|---|---|---|
| $\lambda_1$ (nm) | $\lambda_2$ (nm) | $c_1$ (cm³/mol) | $c_2$(cm³/mol) | $d_3$ (cm³/(mol·μm²)) |
| 80.6 | 206.7 | 27.0630 | 1.2408 | -0.0858 |
| $R_{alkyl}(\lambda)$ | | | | |
| $\lambda_1$ (nm) | $\lambda_2$ (nm) | $c_1$ (cm³/mol) | $c_2$(cm³/mol) | $d_3$ (cm³/(mol·μm²)) |
| 84.6 | 183.5 | 4.5162 | 0.0883 | 0 |

Table 3. Coefficients from the respective fit of $R_0(\lambda)$ and b $R_{\text{alkyl}}(\lambda)$ to the three-resonance Sellmeier models given by equation 8.

The fitting coefficients in Table 3 provide valuable information about how each contribution, $R_0$ and $R_{alkyl}$, influences the dispersion spectra of the family of ILs. The strengths of the resonances are very different for $R_0$ and $R_{alkyl}$ as the strengths associated to $R_0$ are much more intense than the associated to $R_{alkyl}$. This difference in the strengths is not strange as $R_0$ contains the contribution of the [mim][BF$_4$] part of the IL to $R$ while $R_{alkyl}$ is the contribution of an isolated CH$_2$ group. Nevertheless, as the influence of $R_{alkyl}$ grows linearly with the number of carbons in the alkyl chain, the exact balance between both set of strengths depends on the alkyl chain length considered. Interestingly, the influence of the IR resonance $d_3$ on $R_{alkyl}$ was found to be negligible, less than 8·10⁻⁵ in the whole spectrum which is below our experimental resolution. In consequence, we forced it to be 0 in the fitting to equation 8 as it is shown in the results in Table 3. The negligible value of the IR resonance means that the dispersion in the IR range is not influenced by the alkyl chain of the imidazolium cation. Furthermore, the position and the strength of the resonances do not coincide with the values obtained for the material dispersion fitting to equation 4. This mismatch arises from the fact that the refractive index was modelled considering the whole IL while in the case of molar refraction we are analyzing the dispersion by splitting the IL in parts. A much more interesting feature to analyze is the reason why the resonance positions for $R_0$ and $R_{alkyl}$ does not coincide. In the case of the resonance $\lambda_1$ the separation of the resonances is only 4.0 nm while in the case of $\lambda_2$, it is much larger, 23.2 nm. In order to explain the differences in the resonance positions, the contribution of the different parts of a reference IL to the absorption spectrum was calculated by *ab initio* methods in accordance with the computational details provided in section 2. The simulated absorption spectrum for the IL used as reference, [C$_4$mim][BF$_4$] is shown in Figure 8.

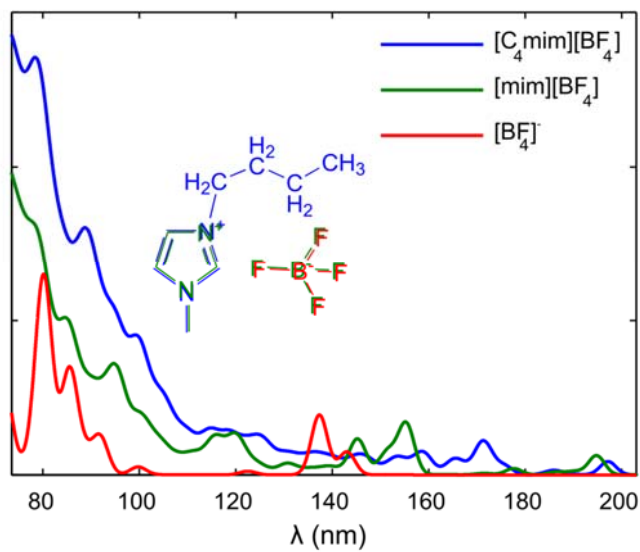

Figure 8. Simulated absorption spectra from different parts of the IL: [C$_4$mim][BF$_4$], [mim][BF$_4$] and [BF$_4$]$^-$.

From Figure 8, it is clear how different is the absorption spectra depending on the part of the IL to be considered. For the resonance $\lambda_1$, two values are found from the fittings, 80.6 nm for $R_0$ and 84,6 nm for $R_{alkyl}$. Analyzing the information of the figure, the [C$_4$mim][BF$_4$] presents two strong absorption peaks around these wavelengths, one at 78.5 nm and the other at 88.9 nm. These peaks present a certain contribution from the [BF$_4$]$^-$ anion and they appear both in the [C$_4$mim][BF$_4$] and in the [mim][BF$_4$] absorption spectra. The fact that the fitting in terms of $R_0$ and $R_{alkyl}$ detects one peak or the other, suggests that there are different influences of the alkyl chain length in both peaks. Specifically, the peak at 88.9 nm is expected to enhance its influence as the alkyl chain is enlarged in comparison with the peak at 78.5 nm. In this regard, the independent fitting of the components of the molar refraction to our modified three-resonance

Sellmeier model is able to discern the influence of the different parts of the IL have on it. This behaviour is more clearly observed for the position of the $\lambda_2$ resonance. In Figure 8 there is a clear absorption peak of [C$_4$mim][BF$_4$] at around 197 nm which is unequivocally associated to the imidazolium ring of the [mim][BF$_4$] part of the IL, and it is reflected in the position of the $\lambda_2$ resonance for $R_0$. In the same spectra, there is also a complex structure in the region from 159 nm to 177 nm which is undoubtedly associated to the butyl alkyl chain. The $\lambda_2$ resonance found at 183.5 nm for $R_{alkyl}$ corresponds with this structure of peaks as in $R_{alkyl}$ the peak of the imidazolium ring at 197 nm is absent. From this point of view, the strategy of separating the contributions of the different parts of the IL to the total $R$ provides a detailed description of the features associated to each part of the IL. Thus, it offers an exciting mechanism for future tailoring the molar refraction dispersion of ILs. With the purpose of obtaining a deeper insight in the behaviour of the resonances with the alkyl chain length, the absorption spectra of some members of the family of ILs studied was calculated. The results of the simulations are shown in Figure 9.

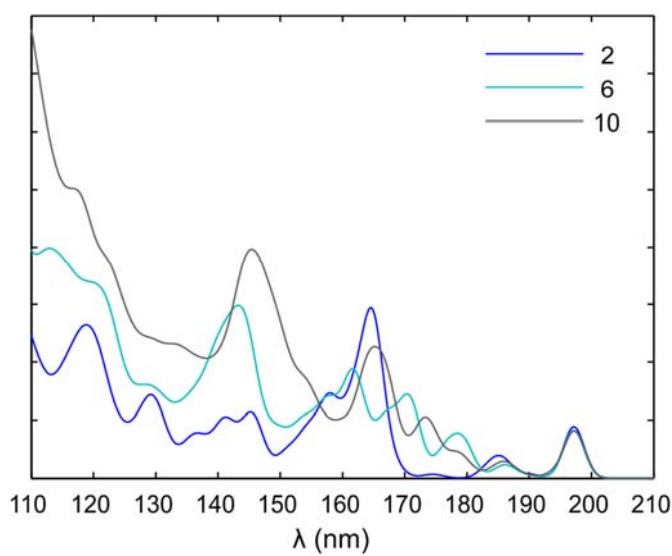

Figure 9. Simulated absorption spectra for the family of ILs [C$_k$mim][BF$_4$] with k=2, 6 and 10.

The absorption spectra of the Figure 9 support some of our previous hypothesis. In first place, the absorption peak placed at 200 nm does not change in shape or position under carbon increase. This behaviour is expected for a resonance which does not depend on the alkyl chain which is the case for this absorption peak as it is related to the basic compound [mim][BF$_4$] and specifically, to its imidazolium ring. Moreover, in the range from 140 nm to 175 nm there is a set of peaks which are strongly dependent both in strength and position on the alkyl chain. This complex structure is probably related to the $\lambda_2$ resonance of $R_{alkyl}$ at 183.5 nm whose position we were not able to exactly reproduce in part due to the erratic behaviour of this structure of peaks with the alkyl chain length. Finally, in the short wavelength region there is a very evident increase of the strength of the absorption peaks with the number of carbons in the alkyl chain which suggest a probable increase of the strength of the $\lambda_1$ resonance of $R_{alkyl}$ with them.

## 4. Prediction of the material dispersion of ILs

It has been stated before that the refractive index of a material can be related to its molar refraction and molar volume through the Lorentz-Lorenz equation, equation 5. Refractive index experimentally depends on temperature and wavelength, however, it has been previously demonstrated that molar refraction is a temperature independent magnitude, at least, within our experimental resolution. From this fact, it can be deduced that the temperature dependence of the refractive index arises only from the implicit temperature dependence of the molar volume $V(T) = M/\rho(T)$ and the refractive index dependence on wavelength comes the dependence of $R(\lambda)$ on the wavelength. Under these considerations, the Lorentz-Lorenz equation describes the refractive index dependence as the product of a pair of totally independent magnitudes: the molar volume only depending on temperature and the molar refraction only depending on wavelength:

$$\frac{n(\lambda,T)^2 - 1}{n(\lambda,T)^2 + 2} = \frac{R(\lambda)}{V(T)} = K_\lambda(\lambda)K_T(T) \tag{11}$$

Equation 11 can be utilized to predict the material dispersion of ILs at whatever temperature if the molar refraction dependence on wavelength and the molar volume dependence on temperature are known.

The molar refraction can be predicted by calculating the electronic polarizability by means of standard *ab initio* calculations as routinely made by other authors[36–39]. However, often these calculations are carried out in the limit of a static field and the information about the wavelength dependence of the electronic polarizability is lost. Nevertheless, there are *ab initio* strategies to predict this wavelength dependence such as the CPKS theory[84]. By means of that theory, the electronic polarizability of the ILs at thirteen different wavelengths, from 300 nm to 1500 nm in 100 nm steps was calculated. On the other hand, molecular volume can be calculated directly through DFT calculations or via density through molecular dynamics (MD) simulations. Both computational approaches provide good results but the small relative deviations obtained from the experimental values [36,85,86] could have a very strong impact in the refractive index prediction. Notwithstanding, nowadays there is available a wide database of experimental densities of ILs and probably, most of them were not optically characterized. For this reason, we decided to take advantage of it and validate our approach for predicting IL material dispersion by obtaining the molar volume through their experimental density. In this work, we use the densities we measured, despite the procedure can be implemented with densities from bibliography without any restriction.

In order to analyze our ability to predict the wavelength dependent behaviour of the electronic polarizability, we compare the molar refraction for our set of ILs calculated by means of the CPKS calculations with the molar refraction obtained from equation 5 using our experimental measurements. We chose to carry out the comparison on this way as experimental and computational contributions can be clearly separated while a direct comparison of refractive indices would mixture both. Figure 10 shows the experimental molar refraction dispersion at 303 K for the seven studied ILs as well as the comparison with the simulated values.

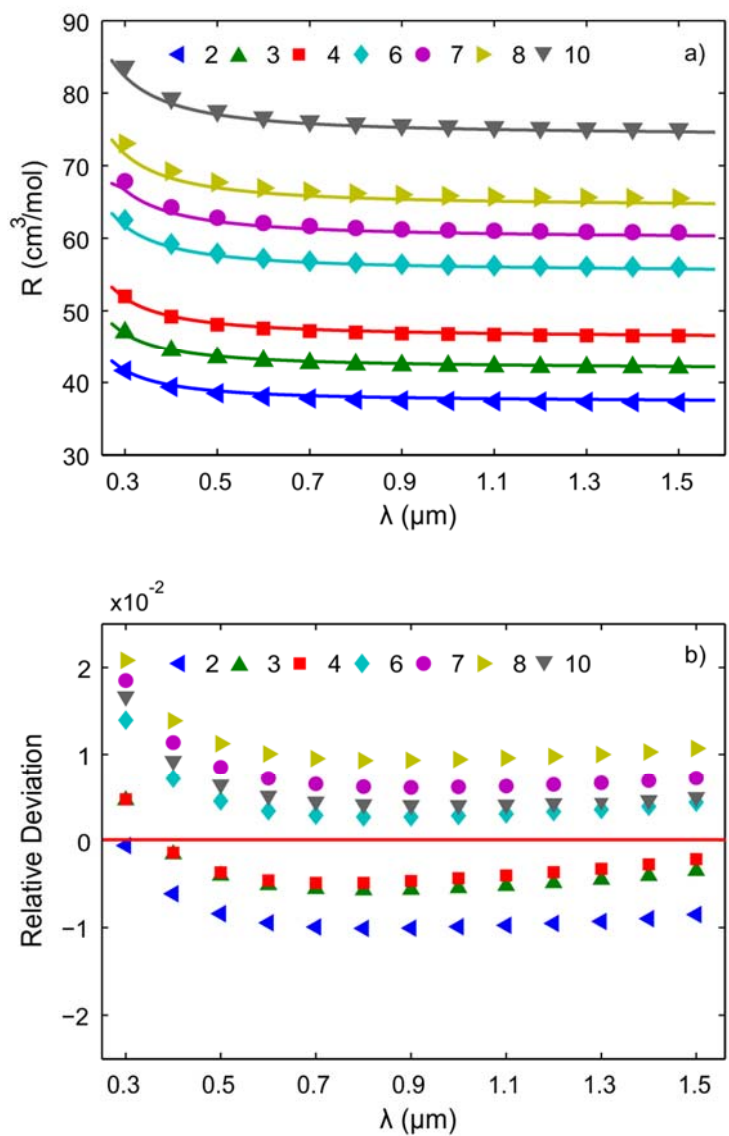

Figure 10. a) Experimental molar refraction dispersion (solid line) and simulated molar refraction dispersion (markers) at 303 K for the 7 analyzed ILs. b) Relative deviation of the simulated data points from the experimental values.

The Figure 10.a), shows how the CPKS calculation is able to approximately reproduce the experimental molar refraction dispersion of this family of ILs with a root mean square (RMS) relative deviation that ranges from $3.9 \cdot 10^{-3}$ (k = 4) to $1.15 \cdot 10^{-2}$ (k = 8). Analyzing the relative deviations between the simulated and experimental values in Figure 10.b), more detailed information can be extracted. In first place, for each compound, the deviation is similar for most of the wavelengths except for the region of shortest wavelengths. In these regions, there is a strong change in the curvature of the deviation for all the compounds. This behaviour could be the consequence of a worst performance of the CPKS method as the simulation goes closer to the position of the resonance previously detected around 200 nm.

On the other hand, the sign of the deviations in almost the whole spectra depends on the alkyl chain length. For compounds with $k \leq 4$, the trend is to underestimate the molar refraction which means that our calculations predict a lower electronic polarizability than the real one. On the contrary, for $k > 4$, the trend is the opposite and there is a clear trend to overestimate the electronic polarizability. In order to obtain a closer insight in this phenomenon, we studied the simulated molar refraction dependence on the alkyl change length at each wavelength. It turned out to be linear, the same behaviour that the experimental molar refraction presents. In consequence, the difference between the experimental and molar refraction, $\Delta R = R_{exp} - R_{sim}$ was calculated and fitted versus the number of carbons in the alkyl chain in our ILs:

$$\Delta R = \Delta R_0 + k \cdot \Delta R_{alkyl} \tag{12}$$

In this equation, $\Delta R_0$ is the difference between the experimental and simulated molar refraction for the limit of having a zero carbons alkyl chain, [mim][BF$_4$]. In this limit, the differences in the molar refraction $\Delta R$ arise from the errors in the prediction of the molar refraction provided by the 1-methylimidazolium cation and the tetrafluoroborate anion. Regarding the $\Delta R_{alkyl}$ term, it describes the contribution to the molar refraction error of each CH$_2$ introduced in the alkyl chain. Figure 11 shows both $\Delta R_0$ and $\Delta R_{alkyl}$ as a function of the wavelength.

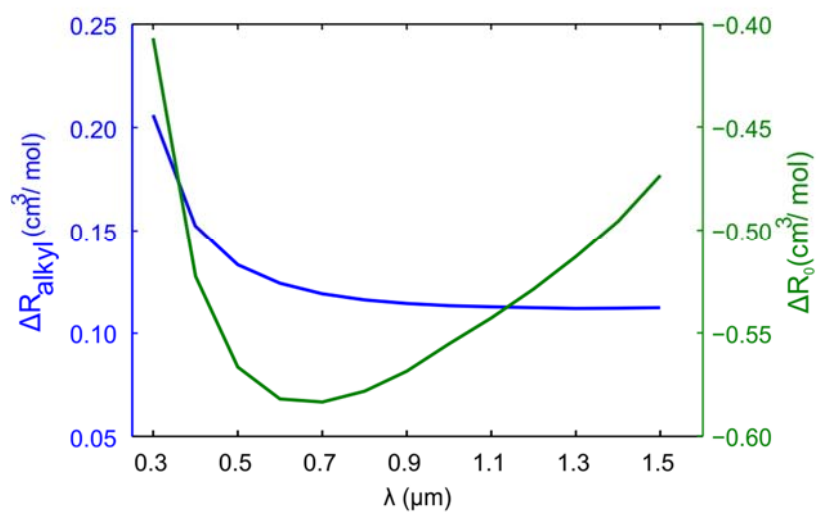

Figure 11. Fitting parameters of a linear fit at different wavelengths of $\Delta R = R_{exp} - R_{sim}$ to the number of carbons, $k$, in the alkyl chain of each compound.

According to Figure 11, both $\Delta R_0$ and $\Delta R_{alkyl}$ present a very different dispersive behaviour and values which can provide valuable information about the performance of our simulation. First of all, it is worth to note that the error in the prediction of the molar refraction comes from an interesting balance between $\Delta R_0$ and $\Delta R_{alkyl}$. The molar refraction of $\Delta R_0$ is always underestimated while the molar refraction of $\Delta R_{alkyl}$ is always overestimated and its influence grows with the length of the alkyl chain. The total error in the molar refraction arises from the combination of these two values. The mean value of $\Delta R_{alkyl}$ in the whole spectral range is 0.13 cm$^3$/mol while the mean value for the $\Delta R_0$ is 0.53 cm$^3$/mol, which is approximately 4 times the error of $\Delta R_{alkyl}$. In consequence, as the error associated to $\Delta R_{alkyl}$ increases with the number of carbons in the alkyl chain, for $k > 4$ the overestimation associated to $\Delta R_{alkyl}$ dominates over the underestimation provided by $\Delta R_0$. This error balance explains the existence of two families of deviations as a function of the alkyl chain of the simulated IL. Regarding the error as a function of the wavelength, the $\Delta R_0$ presents a very dispersive behaviour having a variation in the whole spectral range of 0.17 cm$^3$/mol. On the other hand, $\Delta R_{alkyl}$ presents a much less dispersive behaviour and its variation in the spectral range is 0.10 cm$^3$/mol, being the most important contribution (0.054 cm$^3$/mol) the step from 400 nm to 300 nm. In consequence, from the spectral point of view, the deviation contribution of the alkyl chain is very achromatic (except for 300 nm) and the main contribution to the chromatic error is provided by the limitations in the calculation of the dispersive response of the [mim][BF$_4$] unit.

In our model, the thermal dependence of the refractive index is totally attributed to the molar volume dependence on temperature. Hence, applying the equation 11 it is possible to predict the refractive index at the same wavelength range we studied the molar refraction for different temperatures. These calculations imply merging the computationally obtained electronic polarizability dispersion with the experimental density data we measured at temperatures from 293 K to 313 K. The results are shown in Figure 12.

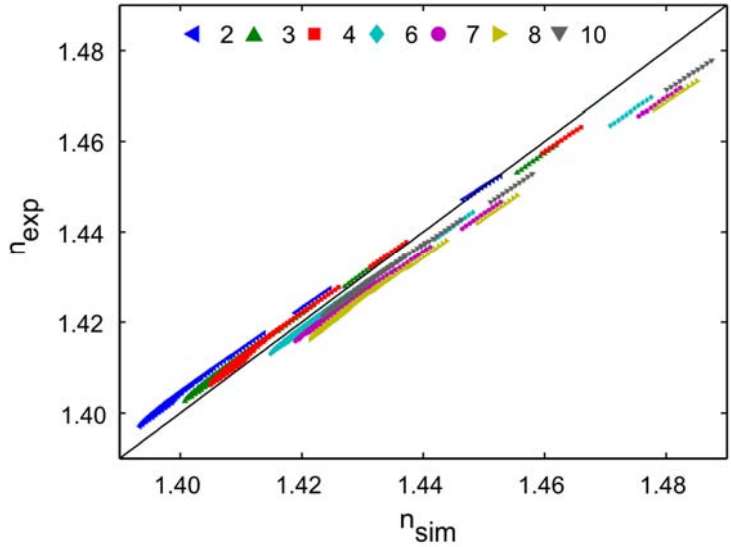

Figure 12. Simulated refractive index, $n_{sim}$, versus experimental refractive index, $n_{exp}$ for the ILs. The shown refractive indices belong to all the simulated wavelengths, from 300 nm to 1500 nm in steps of 100 nm, and temperatures, from 293 K to 313 K in steps of 2 K.

In Figure 12 the refractive index predicted for the seven ILs at all the combinations of temperature, 293 K to 313 K in steps of 2 K and wavelength, 300 nm to 1500 nm in steps of 100 nm, are shown. The RMS relative deviation of the simulated data with regard to the experimental one considering all the wavelengths and temperatures is less than $4 \cdot 10^{-3}$. The obtained relative deviation is very good especially if we consider the experimental relative standard deviation of refractive indices of this family, $\sigma_{exp} = 1.14 \cdot 10^{-3}$ from the fitting results of Figure 2. For each specific compound, the difference between both predicted and experimental refractive indices is uniform in the considered thermal and spectral ranges with exception of the region at the shortest simulated wavelength, 300 nm. This behaviour implies that the model reflects the temperature effect on refractive index on a proper way and that the main deviations arise from the limitations in the prediction of the dispersive component of the refractive index, especially at the shortest wavelengths.

Comparing the performance of our refractive index prediction approach with other published approaches[31,33,87], we obtain similar or better relative deviations despite using a much simpler strategy. However, there are some differences that are worth to note. These previous models are only able to predict the refractive index at the sodium D line while our approach provides refractive index at a wide spectral range and specific temperature. On the other hand, these previous works were tested over a much wider amount of ILs while we restricted our analysis only to the $[C_kmim][BF_4]$ family as a natural extension for understanding the relation of the dispersion properties of these ILs with their structure.

## 5. Conlusion

The material dispersion of the $[C_kmim][BF_4]$ family of ILs with $k=$ 2, 3, 4, 6, 7, 8 and 10 was experimentally measured at several temperatures in a broad spectral range covering from 300 nm to 1550 nm by means of Refractive Index Spectroscopy by Broadband Interferometry. It was demonstrated that in such a wide spectral range a single-resonance Sellmeier dispersion formula is not enough to describe the behaviour of the material dispersion. For this reason, a modified three-resonance Sellmeier dispersion formula was proposed to describe the dispersive behaviour of the ILs and to analyze the influence of the temperature and the alkyl chain in the dispersion. In the employed model, two resonances are placed in the UV region while a third one is placed in the IR. The two UV resonances present a strong influence in the dispersion while the IR resonance only presents a very limited contribution to it. The temperature does not affect the position of the resonances but only affects the strength of the resonance placed in the shortest UV wavelength. The dependence of this resonance strength on the temperature is clearly linear and its magnitude decreases as temperature rises. This negative slope of the resonance strength with respect the temperature is tightly related to the thermo-optical coefficient of this family of ILs. Regarding the dispersion dependence on the alkyl chain length, it was analyzed in terms of the molar refraction. The molar refraction was split into two contributions, one from the variable alkyl chain and the other from the rest of the IL, the 1-methylimidazolium tetrafluoroborate. The dispersion of each contribution was fitted to the same Sellmeier model and different resonances were obtained from each part of the IL. The position of the resonances found for each part of the IL was compared with simulated absorption spectra for this set of ILs and a good agreement was found. It implies that each part of the molecule contributes in a different way to the dispersive behaviour of the total IL. Thus, the dispersive behaviour of an IL can be tailored by properly choosing its structure. Finally, a semi-empirical model was proposed for the prediction of the material dispersion of ILs in the same broad spectral range and temperatures as experimental measurements. This model assumes that the material dispersion is well described by the Lorentz-Lorenz formula and that the refractive index can be described as the product of two independent magnitudes with a common dependence on the alkyl chain length of the compound: the molar volume which depends on temperature and the electronic polarizability being which depends on wavelength. Note that this description implies neglecting the very weak dependence of the electronic polarizability on temperature. The temperature dependence of the molar volume is obtained from experimentally measured densities despite it could be perfectly obtained from the wide density bibliography available for ILs. The wavelength dependence on the electronic polarizability was simulated by DFT using the CPKS strategy. The predicted material dispersion presented a very good agreement for the 7 studied ILs in all the temperatures and wavelengths, being the RMS relative deviation less than $4 \cdot 10^{-3}$. Further analysis of the model provided a detailed description about its performance at each part of the ILs. The main cause of the deviations is not the temperature description in the model but the limitations in the prediction of the electronic polarizability dispersion. In consequence, enhanced performance of the model could be obtained employing a higher level of theory in the electronic polarizability dispersion calculations.

## 6. Supporting information

Table S1. Experimental densities in $g/cm^3$ of the selected ILs at 11 different temperatures from 293 to 313 K each 5 K.

Table S2. Experimental refractive index at selected wavelengths at T=293 K.

Table S3. Experimental refractive index at selected wavelengths at T=303 K.

Table S4. Experimental refractive index at selected wavelengths at T=313 K.

## 7. Acknowledgements

This work was supported by Ministerio de Economía y Competitividad (MINECO) and FEDER Program through the projects (MAT2017-89239-C2-1-P, MAT2017-89239-C2-2-P); Xunta de Galicia and FEDER (AGRU 2015/11 and GRC ED431C 2016/001, ED431D 2017/06, ED431E2018/08). C. D. R. F. thanks the support of Xunta de Galicia through the grant ED481A-2018/032.

# Supporting Information

# An Experimental and Computational Study on Material Dispersion of 1-Alkyl-3-Methylimidazolium Tetrafluoroborate Ionic Liquids

Carlos Damián Rodríguez Fernández[a], Yago Arosa[a], Bilal Algnamat[a,b], Elena López Lago[a*], Raúl de la Fuente[a]

[a] *Nanomateriais, Fotónica e Materia Branda (NaFoMat), Departamento de Física Aplicada e Departamento de Física de Partículas, Universidade de Santiago de Compostela, Campus Vida, E-15782 Santiago de Compostela, Spain*

[b] *Department of Physics, College of Science, Al-Hussein Bin Talal University, Ma'an, Jordan*

*\*Corresponding author: elena.lopez.lago@usc.es*

Table S1. Experimental densities in g/cm$^3$ of the selected ILs at 11 different temperatures.

| $\rho$ (g/cm$^3$) | | | | | | |
|---|---|---|---|---|---|---|
| T (K) | [C$_2$Mim] [BF$_4$] | [C$_3$Mim] [BF$_4$] | [C$_4$Mim] [BF$_4$] | [C$_6$Mim] [BF$_4$] | [C$_7$Mim] [BF$_4$] | [C$_8$Mim] [BF$_4$] | [C$_{10}$Mim] [BF$_4$] |
| 293 | 1.284358 | 1.238723 | 1.204041 | 1.148754 | 1.126412 | 1.105994 | 1.066633 |
| 295 | 1.282830 | 1.237252 | 1.202615 | 1.147344 | 1.125030 | 1.104628 | 1.065257 |
| 297 | 1.281303 | 1.235780 | 1.201193 | 1.145930 | 1.123640 | 1.103250 | 1.063879 |
| 299 | 1.279776 | 1.234306 | 1.199769 | 1.144513 | 1.122252 | 1.101871 | 1.062504 |
| 301 | 1.278257 | 1.232833 | 1.198353 | 1.143160 | 1.120861 | 1.100490 | 1.061129 |
| 303 | 1.276741 | 1.231365 | 1.196937 | 1.141802 | 1.119475 | 1.099112 | 1.059756 |
| 305 | 1.275226 | 1.229899 | 1.195519 | 1.140441 | 1.118134 | 1.097730 | 1.058382 |
| 307 | 1.273712 | 1.228434 | 1.194140 | 1.139080 | 1.116799 | 1.096369 | 1.057007 |
| 309 | 1.272196 | 1.226969 | 1.192694 | 1.137711 | 1.115456 | 1.095046 | 1.055636 |
| 311 | 1.270689 | 1.225511 | 1.191276 | 1.136346 | 1.114119 | 1.093716 | 1.054265 |
| 313 | 1.269186 | 1.224050 | 1.189870 | 1.134985 | 1.112780 | 1.092389 | 1.052892 |

Table S2. Experimental refractive index at selected wavelengths at T=293 K (the letters correspond to the denomination of the Fraunhofer lines)

| | P (336.1 nm) | F (486.1 nm) | D (589.3 nm) | He:Ne (633 nm) | Z (822.7 nm) | (1470 nm) |
|---|---|---|---|---|---|---|
| $[C_2Mim][BF_4]$ | 1.4398 | 1.4185 | 1.4129 | 1.4114 | 1.4073 | 1.4025 |
| $[C_3Mim][BF_4]$ | 1.4465 | 1.4247 | 1.4190 | 1.4173 | 1.4130 | 1.4077 |
| $[C_4Mim][BF_4]$ | 1.4505 | 1.4287 | 1.4229 | 1.4213 | 1.4168 | 1.4114 |
| $[C_6Mim][BF_4]$ | 1.4571 | 1.4353 | 1.4296 | 1.4280 | 1.4236 | 1.4184 |
| $[C_7Mim][BF_4]$ | 1.4594 | 1.4377 | 1.4319 | 1.4303 | 1.4261 | 1.4210 |
| $[C_8Mim][BF_4]$ | 1.4611 | 1.4394 | 1.4337 | 1.4321 | 1.4279 | 1.4235 |
| $[C_{10}Mim][BF_4]$ | 1.4655 | 1.4438 | 1.4380 | 1.4364 | 1.4321 | 1.4274 |

Table S3. Experimental refractive index at selected wavelengths at T=303 K.

| | P (336.1 nm) | F (486.1 nm) | D (589.3 nm) | He:Ne (633 nm) | Z (822.7 nm) | (1470 nm) |
|---|---|---|---|---|---|---|
| $[C_2Mim][BF_4]$ | 1.4372 | 1.4159 | 1.4103 | 1.4088 | 1.4047 | 1.3999 |
| $[C_3Mim][BF_4]$ | 1.4435 | 1.4219 | 1.4162 | 1.4146 | 1.4104 | 1.4052 |
| $[C_4Mim][BF_4]$ | 1.4477 | 1.4260 | 1.4202 | 1.4186 | 1.4143 | 1.4089 |
| $[C_6Mim][BF_4]$ | 1.4541 | 1.4324 | 1.4267 | 1.4251 | 1.4209 | 1.4158 |
| $[C_7Mim][BF_4]$ | 1.4563 | 1.4348 | 1.4291 | 1.4275 | 1.4233 | 1.4185 |
| $[C_8Mim][BF_4]$ | 1.4581 | 1.4365 | 1.4308 | 1.4292 | 1.4250 | 1.4204 |
| $[C_{10}Mim][BF_4]$ | 1.4623 | 1.4407 | 1.4350 | 1.4334 | 1.4292 | 1.4244 |

Table S4. Experimental refractive index at selected wavelengths at T=313 K.

| | P (336.1 nm) | F (486.1 nm) | D (589.3 nm) | He:Ne (633 nm) | Z (822.7 nm) | (1470 nm) |
|---|---|---|---|---|---|---|
| $[C_2Mim][BF_4]$ | 1.4346 | 1.4133 | 1.4078 | 1.4062 | 1.4021 | 1.3972 |
| $[C_3Mim][BF_4]$ | 1.4405 | 1.4190 | 1.4135 | 1.4119 | 1.4078 | 1.4028 |
| $[C_4Mim][BF_4]$ | 1.4449 | 1.4233 | 1.4176 | 1.4160 | 1.4117 | 1.4064 |
| $[C_6Mim][BF_4]$ | 1.4511 | 1.4295 | 1.4239 | 1.4223 | 1.4181 | 1.4132 |
| $[C_7Mim][BF_4]$ | 1.4533 | 1.4319 | 1.4263 | 1.4247 | 1.4206 | 1.4159 |
| $[C_8Mim][BF_4]$ | 1.4551 | 1.4335 | 1.4279 | 1.4263 | 1.4221 | 1.4174 |
| $[C_{10}Mim][BF_4]$ | 1.4592 | 1.4377 | 1.4320 | 1.4304 | 1.4262 | 1.4214 |